\documentstyle[12pt]{article}
 \newcommand{\be}{\begin{equation}}
 \newcommand{\ee}{\end{equation}}
 \newcommand{\br}{\begin{equation}\begin{array}{lll}}
 \newcommand{\er}{\end{array}\end{equation}}
 \newcommand{\ba}{\begin{eqnarray}}
 \newcommand{\ea}{\end{eqnarray}} 
 \newcommand{\ban}{\begin{eqnarray*}}
 \newcommand{\ean}{\end{eqnarray*}} 
 \def\ds{\displaystyle}
 \def\Schw{Sch\-warz\-sch\-ild}
 \def\Sh{Shan\-mu\-ga\-dha\-san}
 
 \def\d{\partial}

 \def\gmn{g_{\mu\nu}}

 \def\real{{\vrule height 1.6ex
 width 0.05em depth 0ex \kern -0.06em {\rm R}}}
 
 \def\sqr#1#2{{\vcenter{\hrule height.#2pt \hbox{\vrule width.#2pt
 height#1pt\kern#1pt \vrule width.#2pt} \hrule height.#2pt}}}
 \def\bar{\overline}

 \def\PRD#1#2#3{{\it Phys.\ Rev.} {\bf D#1}, #2 (#3)}
 \def\PRL#1#2#3{{\it Phys.\ Rev.\ Lett.} \ {\bf #1}, #2 (#3)}
 \def\NPB#1#2#3{{\it Nucl.\ Phys.} {\bf B#1}, #2 (#3)}

 \def\PLB#1#2#3{{\it Phys.\ Lett.} {\bf B#1}, #2 (#3)}

 \def\JMP#1#2#3{{\it J.\ Math.\ Phys.} {\bf #1}, #2 (#3)}

 \def\IJMPA#1#2#3{{\it Int.\ J.\ Mod.\ Phys.} {\bf A#1}, #2 (#3)}
 \def\IJMPD#1#2#3{{\it Int.\ J.\ Mod.\ Phys.} {\bf D#1}, #2 (#3)}
 \def\MPLA#1#2#3{{\it Mod.\ Phys.\ Lett.} {\bf A#1}, #2 (#3)}

 \def\ANP#1#2#3{{\it Ann.\ Physics (N.Y.)} {\bf #1}, #2 (#3)}
 
\begin{document}
 \title{{\bf Static Quantization of Two-dimensional\\
 Dilaton Gravity and Black Holes}}
 \author{Marco Cavagli\`a\thanks{E-mail: cavaglia@aei-potsdam.mpg.de}\\
 \small\sl Max-Planck-Institut f\"ur Gravitationsphysik,\\
 \small\sl Albert-Einstein-Institut\\
 \small\sl Am M{\"u}hlenberg 5, D-14476 Golm, Germany.\\\\
         Vittorio de Alfaro\thanks{E-mail: vda@to.infn.it}\\
 \small\sl Dipartimento di Fisica Teorica dell'Universit\`a di Torino,\\ 
 \small\sl Via Giuria 1, I-10125 Torino, Italy;\\ 
 \small\sl INFN, Sezione di Torino, Italy. }
 \date{}
 \maketitle
 \begin{abstract}
 
Two-dimensional matterless dilaton gravity is a topological theory and can be
classically reduced to a (0+1)-dimensional theory with a finite number of
degrees of freedom. If quantization is performed, a simple gauge invariant
quantum mechanics is obtained. The properties of the gauge invariant operators
and of the Hilbert space of physical states can be determined. In particular,
for $N$-dimensional pure gravity with $(N-2)$-dimensional spherical symmetry,
the square of the ADM mass operator is self-adjoint, not the mass itself.

 \end{abstract}

\section{Introduction}
Recently, the investigation of lower-dimensional gravitational models
\cite{web} has received a large amount of attention because of its relation to
higher-dimensional gravity, integrable systems and black hole physics.

The aim of this paper is to discuss some interesting classical and quantum
properties of the (1+1)-dimensional model of dilaton gravity
\cite{LGK}-\cite{BJL}
\be
S_{DG}=\int d^2x\, \sqrt{-h}\, \left(\phi R^{(2)}(h)+W(\phi) \right)\,,
\label{action} 
\ee
where $\phi$ is the dilaton field, $W(\phi)$ is the dilaton potential, and
$R^{(2)}(h)$ is the two-dimensional Ricci scalar. (Here and throughout the
paper we use Landau-Lifshits conventions \cite{LL} for the Ricci scalar and
natural units.)  

Theories of the form (\ref{action}) may arise from dimensional reduction of
higher dimensional gravity in presence of symmetries. A remarkable example is
dimensional reduction of spherically symmetric vacuum Einstein-Hilbert gravity
in $N>2$ dimensions
\be
S^{(N)}={1\over 16\pi l_{pl}^{N-2}}\int d^Ny\,\sqrt{-g}\,R^{(N)}(g)\,.\label{EH}
\ee
Using for the $N$-dimensional metric the  $(N-2)$-spherically symmetric 
ansatz ($\mu,\nu=0,1$)
\be
ds^2_N=[\phi(x)]^{-(N-3)/(N-2)}\,h_{\mu\nu}(x)\,dx^\mu dx^\nu +
[\gamma\phi(x)]^{2/(N-2)}\,d\Omega^2_{N-2}\,,~~~~~\phi>0\,,
\label{metric-N}
\ee
Eq.\ (\ref{EH}) can be cast in the form (\ref{action}) where
\be
W(\phi)=(N-2)(N-3)(\gamma^2\phi)^{-1/(N-2)}\,,  
\label{W}
\ee
$\gamma=16\pi \, l_{pl}^{N-2}/V_{N-2}$ and $V_{N-2}=2\pi^{(N-1)/2}/\Gamma((N-1)/2)$
is the volume of the $(N-2)$-dimensional unit sphere $\Omega^2_{N-2}$. Note
that we have neglected the surface term
\be
\partial S=-{N-1\over N-2}\int d^2x\,\sqrt{-h}\,\nabla^2\phi\,.
\ee

Dilaton gravity theories defined by Eq.\ (\ref{action}) are classically
integrable \cite{Filippov}. A number of statements are equivalent:
\begin{enumerate}
\item[i)] Dilaton gravity theories reduce to (0+1)-dimensional theories. Any
solution can be represented (in suitable coordinates) as function of only one
coordinate, a property sometimes referred to as ``staticity property'' 
though the Killing vector is not timelike and orthogonal to a spacelike
hypersurface on the entire manifold. 
\item[ii)] Dilaton gravity theories are topological theories.
\item[iii)] A locally conserved (gauge invariant) quantity exists (for
spherically symmetric gravity this coincides with the ADM \cite{MTW}
mass of the system) and defines the horizon(s).
\item[iv)] The only gauge invariant quantities are the locally conserved
quantity and its conjugate momentum.
\end{enumerate} 
These properties can be easily proved using the dilaton and the locally
conserved quantity (and their conjugates) as coordinates in the phase space
(``geometrodynamical variables'' \cite{Cavaglia}). 

A further, conjectured property that should be mentioned is the equivalence of
dilaton gravity theories to a couple of D'Alembert (free) fields (plus a
single degree of freedom). This is proved through the explicit identification
of a canonical transformation in the case of a potential of the form
\cite{Filippov}
\be 
W(\phi) = a e^{c\phi} + b e^{-d\phi}\,.
\label{free-explicit}
\ee
where $a,b,c,d$ are constant parameters. However, in the general case the proof
of existence of a free field representation \cite{Navarroetal} is not
sufficient for the explicit construction of the canonical transformation.

These classical properties are  an essential guideline in choosing the
quantization scheme so as to preserve them. This is evident in the scheme that
has been worked out explicitly in the case of the CGHS model
\cite{BJL,KRV,CDFPhL} using free fields, and in the general case using
geometrodynamical variables \cite{Cavaglia}.

Thanks to the staticity property, dilaton gravity can be quantized according to
two alternative approaches \cite{Cavaproc}. The first approach is implemented
by the explicit reduction of the (1+1)-dimensional system to the couple of
gauge invariant variables (the locally conserved quantity and its conjugate
momentum) (see e.g.\ \cite{Cavaglia,Kuchar,Varadarajan}). In the second
approach the system is classically reduced to a gauge (0+1)-dimensional problem
and then quantized, leading to a quantum mechanical theory. In both cases the
ensuing quantum theory is described by a Hilbert space spanned by the
eigenstates of the quantum operator corresponding to the gauge invariant
quantity (alternatively, its conjugate) and the two approaches formally lead to
the same Hilbert space (``Quantum Birkhoff Theorem''
\cite{Cavaglia,CDFPhL,Cavaproc}).

Although the two approaches are formally equivalent, the second method has the
advantage that canonical quantities are explicitly represented as differential
operators, the Hilbert space is explicitly defined, and the Hermiticity
properties of operators are controlled. This harvest of results is typical of
the quantum mechanical approach and cannot be obtained by the (1+1)-dimensional
method, neither by the direct reduction to the couple of gauge invariant
quantities nor by reduction to free fields. The quantum mechanical approach
has been worked in detail for the \Schw\ black hole system \cite{bh1,bh2}. In
this case, a remarkable result of the (0+1)-dimensional method is that the
square of the ADM mass, not the ADM mass itself, is self-adjoint. 

In this paper we are interested in extending the quantum mechanical treatment
originally developed for the \Schw\ black hole to the entire class of dilaton
gravity models described by Eq.\ (\ref{action}). Our purpose is to show that
the self-adjointness properties of the quantum operators corresponding to the
gauge invariant quantities depend on the particular model under consideration.
In particular, we will show that the non self-adjointness of  the ADM mass is
not a general property of dilaton gravity theories: it holds for pure dilaton
theories that correspond to hyperspherically symmetric metrics and does not
depend on the dimension of space time.

These results suggest that the role of the \Schw\ black hole mass in gravity
can be clarified by the simple quantum procedure model considered in this paper
and that the root of positivity of the ADM mass in the \Schw\ black hole
geometry can be found in its quantum realization. 

The outline of the paper is as follows. In the next section we present the
classical canonical theory. In Sect.\ 3 we focus attention on the models
derived from dimensional reduction of spherically symmetric gravity. Finally,
in Sect.\ 4 we deal with the quantum theory and discuss the self-adjointness of
the relevant operators. 
\section{General Canonical Theory}
We parametrize the two-dimensional metric $h_{\mu\nu}$ as \cite{BJL}
\be
\gmn=\rho\left(\matrix{\alpha^2-\beta^2&\beta\cr
\beta&-1\cr}\right)\,.
\label{metric} 
\ee
Here $\alpha>0$ and $\beta>0$ play the role of the lapse function and of the
shift vector respectively; $\rho(x_0,x_1)$ represents the dynamical
gravitational degree of freedom. The coordinates $x_0$, $x_1$ are both defined
on $\real$. 

The Hamiltonian form of Eq.\ (\ref{action}) is 

\be 
S=\int d^2x\,
\left[\dot\rho\pi_{\rho}+\dot\phi\pi_\phi-\alpha {\cal H}-
\beta{\cal P}\right]\,,
\label{action-canonical}
\ee
where the super-Hamiltonian and super-momentum are
\ba
&&{\cal H}=\rho\pi_\rho\pi_\phi+{\rho'\over\rho}\phi'-2\phi''-\rho\,
W(\phi)\,,\label{superH}\\
&&{\cal P}=-\phi'\pi_\phi+\rho'\pi_\rho+2\rho\pi_\rho'\,,\label{superP}
\ea
respectively. (We neglect boundary terms as they are irrelevant to the
following discussion. See e.g.\ \cite{Cavaglia, KRV, Kuchar} and references
therein.) 

According to the statement i) of Sect.\ 1 any classical solution, in suitable
coordinates, can be written as a function of a single coordinate
\cite{Filippov}. Thus the problem is reduced to a problem of finite degrees of
freedom. In the canonical framework we can impose the staticity condition 
by requiring that both the metric and the dilaton and their momenta depend on a
single variable. Setting
\be
\alpha\equiv\alpha(x_0)\,,~~~\rho\equiv\rho(x_0)\,,~~~
\phi\equiv\phi(x_0)\,, ~~~\pi_{\rho}\equiv\pi_{\rho}(x_0)\,,
~~~\pi_{\phi}\equiv\pi_{\phi}(x_0),
\ee
the action (\ref{action-canonical}) is cast in the form 
\be
S=\int dx_0\,\left[\dot\rho\pi_\rho+\dot\phi\pi_\phi-l(x_0)H\right]\,,
\label{action1d}
\ee
where
\be
l(x_0)\equiv\alpha\rho(x_0)
\label{lagrange}
\ee
is a Lagrange multiplier enforcing the constraint $H=0$. The super-momentum
constraint defined in Eq.\ (\ref{superP}) vanishes identically and $H$
corresponds to the (0+1)-dimensional slice of the super-Hamiltonian in Eq.\
(\ref{superH}). This is given by
\be
H=\pi_\rho\pi_\phi-W(\phi)\,.
\label{hamil1d}
\ee
Two remarks are in order. Firstly, the action (\ref{action1d}) should be
interpreted as a density action in the coordinate $x_1$, i.e.\ $[S]=[{\rm
length}^{-1}]$. Alternatively, the coordinate $x_1$ can be made compact. In
this case we set for simplicity ${\rm Vol}\,(x_1)=1$. The second remark
concerns the definition of the Lagrange multiplier $l(x_0)$. Equation
(\ref{lagrange}) is meaningful only if $\rho$ has definite sign. Indeed, the
gauge evolution parameter is a monotonic increasing function in $x_0$ on all
trajectories provided that the Lagrange multiplier has definite sign -- at
least on the constraint surface. (Possibly, some simple zeroes may be harmless
but one cannot make any general statement about this.) Therefore, in the
following we will restrict attention on strictly positive values of $\rho$.
(The discussion for $\rho<0$ is analogous and leads to the same canonical
equations, the only difference being the overall sign of the gauge parameter.)
This condition can be lifted if one requires the continuity of the canonical
variables at any space time point. Indeed, on the constraint surface the
equation $\rho=0$ defines the horizon(s) of the two-dimensional metric
(\ref{metric}) -- see Eq.\ (\ref{sol-canonical}) below. So, by requiring the
continuity of the canonical variables across the horizon(s) the dynamics
generated by Eqs.\ (\ref{action1d}) and (\ref{hamil1d}) holds for any value of
$\rho$.

The gauge equations or, alternatively, the (unconstrained) equations of motion
can be easily integrated. The result is
\br
\ds
\phi={\tau\over 2I}\,,
&&\pi_\phi=2I[H+W(\phi)]\,,\\\\
\rho=4I^2[N(\phi)+H\phi-J/2]\,,
&&\ds \pi_\rho={1\over 2I}\,,
\label{sol-canonical}
\er
where $I$ and $J$ are two gauge invariant quantities, $N(\phi)=\int d\phi
\,W(\phi)$, and 
\be
\tau(x_0)=\int_{0}^{x_0} d\bar x_0\, l(\bar x_0)
\ee
is the gauge parameter. The gauge invariant quantities $I$ and $J$ can be
written as functions in the phase space:
\ba
&&I={1\over 2\pi_\rho}\,,\\
&&J=2[N(\phi)+H\phi-\rho\pi_\rho^2]\,.
\ea
Clearly, since $I$ and $J$ are gauge invariant, their Poisson brackets with $H$
must vanish (at least weakly). By direct calculation one can prove that
actually the Poisson brackets vanish strongly. Moreover one also finds
$[J,I]_P=1$. By completing the triplet $I$, $J$, $H$ by
\be
Y={\phi\over\pi_\rho}\,,~~~~[Y,H]_P=1\,,
\ee
one obtains a maximal set of gauge invariant canonical variables
\cite{LeviCivita}, often referred as ``\Sh'' variables \cite{Shan}.  Using the
\Sh\ variables the action (\ref{action1d}) assumes the simpler form
\be
S = \int dx_0\, \left[\dot JI+\dot YH-l(x_0)H\right]\,.
\label{action1d-sh}
\ee
For sake of completeness let us note that the gauge invariant quantity $J$ is
related to the (0+1)-dimensional slice of  the conserved local quantity $M$ 
 as defined by \cite{Filippov} (see also \cite{LGK,LK}) 
\be
M=\int_0^{\phi} W(\bar\phi) d\bar\phi-
{\rho^2\pi_{\rho}^2-\phi'^2\over\rho}\,.
\ee
By a simple algebra one can prove that $M_{|0+1}=J/2-H\phi$. We will see later
that in the spherically symmetric reduced models $J$ coincides (apart from some
numerical factors) with the ADM mass of the system. 

Let us conclude this section by an interesting remark concerning the support of
the gauge invariant quantity $I$. From the first Eq.\ (\ref{sol-canonical}) we
see that $I$ has the sign of $\phi$. Indeed, since $l(x_0)>0$, we can take
$\tau$ positive. (In quantum mechanics one never gets the Feynman propagator
without this positivity restriction, see e.g.\ \cite{Teitelboim,BOL}.) 
This property will be essential in the following.
\section{Spherically Symmetric Gravity}
We have mentioned in the introduction that $(N-2)$-spherically symmetric
gravity in $N$ dimensions can be described by Eq.\ (\ref{action}) where the
dilaton potential is given by Eq.\ (\ref{W}). The connection with
$N$-dimensional spherically symmetric gravity in the standard \Schw\ form can
be better exploited using the ``\Schw-like'' canonical variables
$(a,\pi_a;b,\pi_b)$ \cite{bh1,bh2} defined by the canonical transformation
\br
\ds\rho=2ab\,,&&\ds\pi_\rho={\pi_a\over 2b}\,,\\
\ds\phi={1\over\gamma}b^{N-2}\,,
&&\ds\pi_\phi={\gamma\over N-2}{b\pi_b-a\pi_a\over b^{N-2}}\,.
\label{canonical-transf}
\er
By defining the new Lagrange multiplier 
\be
\bar l(x_0)={\gamma\over N-2}{1\over b^{N-3}}l(x_0)\,,
\ee
the (0+1)-dimensional action (\ref{action1d}) becomes
\be
S=\int dx_0\,\left[\dot a\pi_a+\dot b\pi_b-\bar l(x_0)\bar H\right]\,,
\label{action1d-2}
\ee
where
\be
\bar H={\pi_a\over 2b^2}(b\pi_b-a\pi_a)-kb^{N-4}\,,~~~~
k=(N-2)^2(N-3)\gamma^{-{N-1\over N-2}}\,.
\label{hamil1d-2}
\ee
Using the new canonical chart the $N$-dimensional metric (\ref{metric-N})
reads
\be
ds^2_N=2\,\gamma^{{N-3\over 
N-2}}b^{4-N}\left[-adt^2+ndr^2\right]+b^2d\Omega^2_{N-2}\,,\label{metric-N2}
\ee
where we have set $\alpha=\sqrt{n/a}$ and $x_0=r$, $x_1=t$. Finally, in terms
of the \Schw\ variables the Lagrange multiplier is
\be
\bar l(r)={2\gamma\over N-2}{\sqrt{an}\over b^{N-4}}\,.
\ee
For sake of completeness we give the (on-shell) expression of the gauge
invariant variables $I$ and $J$ in terms of the new canonical variables:
\ba
&&I_{|\bar H=0}={b\over\pi_a}\,,\label{I-new}\\
&&J_{|\bar H=0}={2(N-2)^3\over\gamma^{(N-1)/(N-2)}}b^{N-3}-\pi_a\pi_b\,.
\label{J-new}
\ea
When $N=4$ Eqs.\ (\ref{hamil1d-2})-(\ref{J-new}) coincide with the
corresponding quantities of \cite{bh1,bh2}.

Inserting the solution (\ref{sol-canonical}) in Eq.\ (\ref{metric-N2}) we have
\be
ds^2=-\tilde I^2\left(1-{\tilde J\over b^{N-3}}-{\tilde H\over b^{N-4}}\right)
dt^2+\left(1-{\tilde J\over b^{N-3}}-{\tilde H\over b^{N-4}}\right)^{-1}db^2+
b^2d\Omega^2_{N-2}\,,
\label{sol-metric}
\ee
where 
\ba
&&\tilde I={2(N-2)\over\gamma^{1/(N-2)}}I_{|\bar H=0}\,,\\\nonumber\\
&&\tilde J={\gamma^{(N-1)/(N-2)}\over 2(N-2)^2}J_{|\bar H=0}\,,\\\nonumber\\
&&\tilde H={\gamma^{(N-1)/(N-2)}\over (N-2)^2}\bar H\,.
\ea
The horizons of the $N$-dimensional black hole are defined on the constraint
shell $\tilde H=0$ by 
\be
b^{N-3}=\tilde J\,.
\ee
The ADM mass on the constraint shell is
\be
M_{\rm ADM}={N-2\over\gamma}\tilde J=
{\gamma^{1/(N-2)}\over 2(N-2)}J_{|\bar H=0}\,.
\ee

Let us discuss the support of the canonical variables and of the gauge
invariant quantities. From Eqs.\ (\ref{metric-N}) and (\ref{W}) we have
$\phi>0$, $\phi=0$ being a singularity of the metric. (Clearly, the same
conclusion is obtained using the \Schw\ canonical variables. Indeed, starting
from the metric (\ref{metric-N2}) we have $b>0$ and the canonical
transformation (\ref{canonical-transf}) implies $\phi>0$.) From the discussion
at the end of the previous section it follows $I>0$. This property will play an
essential role in the quantization of the system. The gauge invariant variable
$J$ (and thus the ADM mass) does not have a definite sign in general. Both
positive and negative masses are allowed. In order to exclude negative ADM
masses in spherically symmetric geometries we have to invoke a further,
physically required, ad hoc principle as it is usually the case, or we have to
look for some mechanism responsible for this property. Curiously, this result
is a bonus of the quantum theory of spherically  symmetric geometries, see next
section. 

\section{Quantization} 
The quantization of the model described in the previous sections leads to a
quantum mechanical system with gauge invariance. Here our treatment follows
closely \cite{bh2}. We implement the quantization by the Dirac method,
quantizing first and fixing the gauge after having solved the Wheeler-de Witt
equation.

The quantization of the system is straightforward in the \Sh\ representation.
Formally, the quantization is achieved by imposing first the commutation
relations
\be
[\hat{J},\hat{I}]=im_{pl}\,,~~~~~~~[\hat{Y},\hat{H}]=im_{pl}\,,
\ee
and then by imposing the constraint as a null operator on the states in the
Hilbert space
\be
\hat{H}\Psi=0\,.
\label{WDW}
\ee
In order to represent the canonical coordinates as differential operators we
must first choose a pair of commuting variables as coordinates in the Hilbert
space and establish the form of the (non-gauge fixed) Hilbert measure $d\mu$.
The measure $d\mu$ is determined by the requirement that it is invariant under
the symmetry transformations of the system, namely under the rigid
transformations generated by a couple of suitable gauge invariant quantities
$F(I,J)$ and $G(I,J)$ and under the gauge transformations generated by $H$. In
this process the support of the canonical variables is essential.

Let us suppose that $I$ and $J$ are defined on the whole real axis. This
happens for instance when the dilaton potential $W(\phi)$ -- see Eq.\ (\ref{W})
-- is a well defined functional of the dilaton for any value of $\phi$. In this
case we can choose $Y$ and $I$ as coordinates in the Hilbert space.
(Alternatively, we might choose $Y$ and $J$, the two representations being
related by a Fourier transform.) Denoting by $y$, $x$, $j$ the (continuous)
eigenvalues of $\hat{Y}$, $\hat{I}$, $\hat{J}$, respectively, the gauge and rigid
invariant measure in the Hilbert space is 
\be
d\mu=dxdy\,.
\label{measure-1}
\ee
The differential representation of the operators is 
\be
\hat{I}=x\,,~~~~\hat{J}=im_{pl}{\d~\over\d x}\,,~~~~\hat{Y}=y\,,~~~~
\hat{H}=-im_{pl}{\d~\over\d y}\,.
\ee
By imposing the quantum constraint (\ref{WDW}) we find that the physical states
do not depend on $y$. A basis in the gauge fixed ($y=const$) Hilbert space --
see \cite{bh2} for details -- is given by the set of eigenstates of $\hat{J}$
with eigenvalue $j$
\be
\Psi_j(x)={1\over \sqrt{2\pi m_{pl}}}e^{-ijx/m_{pl}}\,.
\ee

Let us now suppose that the support of $\phi$ does not coincide with the entire
real axis and consider for simplicity $\phi\in\real^{+}$. We have seen in the
previous section that models describing spherically symmetric Einstein gravity
in $N$-dimensions belong to this class. In this case $I>0$ from (15) and we
cannot use the rigid symmetry generated by $J$ to fix the Hilbert measure since
it changes the sign of $I$. Following \cite{bh1,bh2} we define the  gauge
invariant ``dilatation'' generator $N=IJ$. (The dilatation operator was also
introduced in \cite{Kastrup}, to avoid negative masses; here and in [14, 15]
this result is obtained as a consequence of the support properties of the
conjugate variable $I$ and of quantization.) It is easy to check that $N$
generates a symmetry that preserves the sign of $I$ and can be used to
determine the Hilbert measure. Imposing the invariance under the rigid
symmetries generated by $N$ and $I$ the Hilbert measure is 
\be
d\mu={dx\over x}dy\,,~~~~~~x>0\,.
\label{measure-2}
\ee
The measure (\ref{measure-2}) implies that the operator $\hat{J}$ is not
self-adjoint being conjugate to a positive definite operator. Indeed,
according to (\ref{measure-2}) the differential representation of $\hat{J}$ is
\be
\hat{J}=im_{pl}\sqrt{x}{\d~\over\d x}{1\over\sqrt{x}}\,.
\ee
Consequently, the eigenstates of $\hat{J}$ are
\be
\Psi_j(x)=c(j)\sqrt{x}e^{-ixj/m_{pl}}\,.
\ee
It is straightforward to verify that $\hat{J}$ is not self-adjoint in the space
defined by (\ref{measure-2}). A self-adjoint operator in the
space (\ref{measure-2}) is rather $\hat{J}^2$. Thus in spherically symmetric
gravity the square of the ADM mass operator, not the ADM mass operator, is
self-adjoint.  

For sake of completeness, let us give the eigenfunctions of $\hat{J}^2$ (with
eigenvalue $j^2$). We have two separate sets ($j>0$)
\ba
&&\Psi_{j^2}^{(1)}(x)\,=\,{1\over\sqrt{\pi jm_{pl}}}\,\sqrt{x}\,\sin(jx/m_{pl})\,,
\label{J2-1}\\
&&\Psi_{j^2}^{(2)}(x)\,=\,{1\over\sqrt{\pi jm_{pl}}}\,\sqrt{x}\,\cos(jx/m_{pl})\,.
\label{J2-2}
\ea
The effect of the non self-adjoint operator $\hat{J}$ is to transform the set
(\ref{J2-1}) into the set (\ref{J2-2}) and viceversa.
\section{Conclusions}
The (0+1)-dimensional (``static'') canonical quantization of two-dimensional
matterless dilaton gravity shows that the self-adjointness of gauge invariant
operators depends on the global properties of the model. In particular, the
gauge invariant operator $\hat{J}$ that identifies the horizon(s) of the metric
may not have a self-adjoint extension. This happens for models describing
spherically symmetric gravity in $N$ dimensions. In this case  $\hat{J}$ is --
apart from a numerical factor -- the gauge invariant operator corresponding to
the ADM mass of the geometry. Consequently, the ADM mass operator is not
self-adjoint. Instead, its square is self-adjoint and  its eigenfunctions can
be defined in the Hilbert space (with positive eigenvalues of course). This
result (obtained in \cite{bh1}  and \cite{bh2} for the Schwarzschild black
hole) may be the key to  dispose of an ad hoc principle to eliminate negative
masses in spherical geometries, since the only admissible operator is the
square of the mass. See also the discussion contained in \cite{Kastrup} where
the use of the operator $\hat{N}$ is advocated as a principle to avoid negative
values of the mass. 

Regardless of the dilaton potential chosen in Eq.\ (\ref{action}), both
$\hat{J}^2$ and $\hat{J}$ (when the latter can be defined) have continuous
spectra. This result is in agreement with the group theoretical quantization of
SO(3)-symmetric four-dimensional gravity via reduction to a SL(2,R)/SO(2)
non-linear sigma model coupled to three-dimensional gravity \cite{Hollmann,
HBM}. 
 
Quantization of the mass can be achieved by changing the boundary  conditions.
For examples of this procedure, we refer to \cite{bh2}  and especially to
\cite{Kastrup} where further references can be found. Let us remark that there
are indications for a discrete mass spectrum to be obtained by inclusion of
matter in the system. Evidence supporting this conjecture can be found in
\cite{BBN} where the quantization of spherically symmetric gravity coupled to a
thin dust shell is derived. 

\section*{Acknowledgments}
M.C.\ is supported by a Human Capital and Mobility grant of the European Union,
contract number ERBFMRX-CT96-0012.

\end{document}